\documentstyle{article}
\input{epsf}
\newcommand{\be}{\begin{equation}}
\newcommand{\ee}{\end{equation}}
\newcommand{\bea}{\begin{eqnarray}}
\newcommand{\eea}{\end{eqnarray}}
\newcommand{\ba}{\begin{array}}
\newcommand{\ea}{\end{array}}

\def\bbox{{\,\lower0.9pt\vbox{\hrule \hbox{\vrule height 0.2 cm
\hskip 0.2 cm \vrule height 0.2 cm}\hrule}\,}}
\newcommand{\dsl}{\pa \kern-0.5em /}

\newcommand{\nn}{\nonumber \\}
\newcommand{\tr}{{\rm tr}\,}

\def\ben{\begin{equation}}
\def\een{\end{equation}}
\def\bena{\begin{eqnarray}}
\def\eena{\end{eqnarray}}

\def\e{\epsilon}
\def\6{\partial}

\def\today{\ifcase\month\or
  January\or February\or March\or April\or May\or June\or
  July\or August\or September\or October\or November\or December\fi
 \space\number\day, \number\year}

\input amssym.def
\input amssym.tex
\font\mybb=msbm10 at 10pt
\def\bb#1{\hbox{\mybb#1}}

\def\bE {\bb{E}}

\def\bft{\mbox{\boldmath $\tau$}}
\def\bfO{\mbox{\boldmath $\Omega$}}
\def\bfn{\mbox{\boldmath $\nabla$}}
\def\bfG{\mbox{\boldmath $\Gamma$}}
\def\tr{{\rm tr}}
\begin{document}

%%%%%%%%%%%%%%%% title page %%%%%%%%%%%%%%%%%%%%%%%%%%%%%%%%%%%%

\begin{titlepage}
\vfill
\begin{flushright}
QMW-PH-00-05\\
DAMTP-2000-68\\
KCL-00-41\\
hep-th/0008221\\
\end{flushright}

%\centerline{\Large \bf {$\frak{D}\frak{R}\frak{A}\frak{F} \frak {T}$}}
%\centerline { \bf \today}

\vfill
\begin{center}
\baselineskip=16pt
{\Large\bf D-brane Solitons in Supersymmetric Sigma-Models}
\vskip 0.3cm
{\large {\sl }}
\vskip 10.mm
{\bf ~Jerome P. Gauntlett$^{*,1}$, ~Rub{\'e}n Portugues$^{\dagger,2}$,
~David Tong$^{\sharp,\flat,3}$\\
 and  ~Paul K. Townsend$^{\dagger,4}$ } \\
%\\[2mm]
\vskip 1cm
%\vfill
{\small
$^*$
  Department of Physics\\
  Queen Mary and Westfield College\\
  Mile End Rd, London E1 4NS, UK\\
}
\vspace{6pt}
{\small
 $^\dagger$
DAMTP\\
Centre for Mathematical Sciences\\
Wilberforce Road\\
Cambridge CB3 0WA, UK\\
}
\vspace{6pt}
{\small
 $^\sharp$
Dept of Mathematics\\
Kings College, The Strand\\
London, WC2R 2LS, UK\\
}
\vspace{6pt}
{\small
 $^\flat$
Physics Department\\
Columbia University\\ 
New York, NY, 10027, USA\\
}
\end{center}
\vfill
\par

\begin{center}
{\bf ABSTRACT}
\end{center}
\begin{quote}

Massive D=4 N=2 supersymmetric sigma models typically admit domain wall
(Q-kink) solutions and string (Q-lump) solutions, both preserving 1/2
supersymmetry. We exhibit a new static 1/4 supersymmetric `kink-lump' solution 
in which a string ends on a wall, and show that it has an effective 
realization as a BIon of the D=4 super DBI-action. It is also shown 
to have a time-dependent Q-kink-lump generalization which reduces to 
the Q-lump in a limit corresponding to infinite BI magnetic field. 
All these 1/4 supersymmetric sigma-model 
solitons are shown to be realized in M-theory as calibrated, or 
`Q-calibrated', M5-branes in an M-monopole background. 
   
\vfill
 \hrule width 5.cm
\vskip 2.mm
{\small
\noindent $^1$ E-mail: j.p.gauntlett@qmw.ac.uk \\
\noindent $^2$ E-mail: r.portugues@damtp.cam.ac.uk \\
\noindent $^3$ E-mail: tong@physics.columbia.edu\\
\noindent $^4$ E-mail: p.k.townsend@damtp.cam.ac.uk \\
}
\end{quote}
\end{titlepage}
%%%%%%%%%%%%%%%%%%%%%%%%%%%%%%%%%%%%%%%%
\setcounter{equation}{0}
\section{Introduction}

Although D-branes are normally defined within perturbative string theory in 
terms of Dirichlet boundary conditions at the endpoints of 
open strings, they may be defined more generally as branes on 
which strings can end. As such, D-branes may occur in field theories. 
An example is provided by the domain walls (alias
2-branes) of MQCD, which were shown in \cite{MQCD} to be surfaces on which MQCD
strings may end. However, the physics of strings and walls in MQCD is quite
different from that of the D2-branes of IIA superstring theory because the
endpoints of MQCD strings are not electric sources for a gauge field on the
wall. Other examples of (non-supersymmetric) field
theory D-branes have been discussed in \cite{CT}, 
although the physics is again rather different from 
that of string theory D-branes. 

A field theory domain wall that is a much closer analogue of the D2-brane of
IIA superstring theory is provided by the kink domain wall of massive
hyper-K\"ahler (HK) sigma-models \cite{AT}. As pointed out in \cite{AT}, the
effective action for the kink domain wall is the $S^1$ reduction of the D=5
supermembrane, and hence dual to a gauge theory. 
This is similar to the relation
between the D=11 supermembrane and the D=10 D2-brane action \cite{pkt}, and the
same arguments used in that case imply that the gauge theory in question is a
supersymmetric one of Dirac-Born-Infeld (DBI) 
type. As in the D=10 case \cite{CM},
this D=4 action admits 1/2 supersymmetric BIon solutions that 
can be interpreted as
strings ending on a membrane. But what are these sigma-model strings? This
is one of several questions that we aim to answer in this paper. Another is
whether there is a 1/4 supersymmetric sigma-model configuration representing a
string ending on a domain wall, as the analogy with superstring D-branes
suggests. Indeed there is, and for simple models it can be found explicitly and
its properties studied in detail. 

Specifically, we shall consider D=4 supersymmetric sigma models with a
`multi-centre' HK target space 4-metric of the form
\be
ds^2 = U d{\bf X} \cdot d{\bf X} + U^{-1} (d\varphi + d{\bf X}\cdot {\bf
A})^2
\ee
where $\bfn \times A = \bfn U$ and $U$ is a `multi-centre' harmonic function.
The only potential term consistent with maximal supersymmetry is proportional
to the norm of the tri-holomorphic Killing vector field 
$\zeta =\partial/\partial \varphi$, and so takes the form
\be\label{pot}
V= {1\over2}\mu^2 U^{-1}
\ee
where $\mu$ is a mass parameter. Introducing a coupling constant $g$
with dimensions of inverse mass, we have the sigma-model  
Lagrangian density 
\be\label{lag}
{\cal L} = -{1\over 2g^2} \left\{\eta^{\mu\nu}\left[ U\partial_\mu{\bf X}\cdot
\partial_\nu {\bf X} + U^{-1}{\cal D}_\mu \varphi {\cal D}_\nu
\varphi\right] + \mu^2 U^{-1}\right\}
\ee
where $\eta$ is the D=4 Minkowski metric (of `mostly plus' signature)
and ${\cal D}\varphi = d\varphi + d{\bf X}\cdot {\bf A}$. 
When $\mu \ne 0$ we have a `massive' sigma model; otherwise it is
massless.  

The massless sigma models typically admit 1/2 supersymmetric `lump'  
solitons \cite{ward} supported by a topological `lump' charge $L$.
These are of course string-like solitons in D=4. Lump-string
configurations also exist in the massive model, with a string
tension that is bounded from below by the lump charge $L$, but
Derrick's theorem implies that the bound is saturated only in the
limit in which the string core has shrunk to zero size, yielding a
singular field configuration. In other words, the massive sigma-model
admits BPS strings that are `fundamental' in the sense that the core 
size vanishes. One might be tempted to ignore these strings on the
grounds that they are singular, but there are various circumstances in
which the singularity is resolved. For example, the singularity can be
removed, and Derrick's theorem evaded, by incorporating
time-dependence. Indeed, there exists a time-dependent non-singular
charged lump-string solution; its cross section is the D=3 Q-lump solution
found by Abraham \cite{Abraham}. These solutions saturate an energy
bound of the form  
\be
E \ge |L| + |Q|\, .
\ee
where $Q$ is the Noether charge associated with the symmetry 
generated by $\zeta$. Although the solution is not static it is
{\sl stationary} in the sense that the energy density is
time-independent, a fact that allows it to preserve some fraction of
supersymmetry. This fraction was not previously determined but we
shall show here that HK Q-lumps are 1/4 supersymmetric.
Massive HK sigma models also admit kinks (static solitons
that interpolate between the minima of the potential) and Q-kinks.   
The Q-kinks are stationary charged kinks that
saturate an energy bound of the form
\be
E \ge \sqrt{|{\bf K}|^2 + Q^2}
\ee
where ${\bf K}$ is a triplet of topological kink charges. 
The Q-kink with $Q=0$ is the static kink. Both kinks and Q-kinks
preserve 1/2 supersymmetry. 

The main result of this paper is a new non-singular {\sl static} 1/4 
supersymmetric soliton which we call the kink-lump. It has a natural
interpretation as a string ending on a domain wall. To see why such 
configurations might be anticipated, we begin by recalling that the 
D=3 Q-lump can be viewed as a closed loop of D=3 Q-kink-string
\cite{AT2}, so the D=4 Q-lump-string can be viewed as a cylindrical
tube of Q-kink domain wall. If this tube is splayed out at one end we
have a (non-static) configuration representing a string ending on a
wall. If we now remove the charge we might expect to end up with a 
static solution of similar type but with the string core supported
against collapse by its attachment to the wall. The kink-lump is just such 
a solution. The size of the string core decreases with
distance from the wall so its shape is more accurately described as a
`spike' than as a `tube'. Nevertheless, the spike has a constant 
energy per unit
length and can therefore be interpreted as a string of fixed
tension. This tension turns out to equal the tension of the singular 
infinite lump-string, but the kink-lump is completely non-singular
because the `spike' shrinks to zero size only at infinite distance
from the wall.

These results are reminiscent of the BIon solution on a D2-brane
\cite{CM}. For example, the endpoint of the BIon string on a D2-brane
is essentially a global vortex with a logarithmically 
infinite energy, which leads to a logarithmic bending of the D2-brane.
We shall show that the kink-lump incorporates the same logarithmic bending
of the kink domain wall. Moreover, the way in which the singular
lump-string is `blown up' into a cylindrical kink domain wall is 
reminiscent of the way that a IIA superstring can be `blown up' into
a cylindrical D2-brane \cite{emp}. However, there is an important difference. 
The BIon is a solution of 
the  field theory governing the fluctuations of the wall, so the wall 
itself is not part of the solution. The BIon spike remains hollow no 
matter how much it shrinks because the width of the wall itself is
assumed to vanish. In contrast, the kink domain wall is part of the
kink-lump solution and it has a definite thickness; as the spike 
shrinks to a size comparable to the thickness of the wall it must
`fill-in' to form a `solid spike'. We shall see explicitly how this 
happens in the kink-lump solution.
The BIon analogy is really more appropriate to an effective description
of the kink-lump as a 1/2 supersymmetric soliton of the effective theory 
governing fluctuations of a kink domain wall because, as mentioned above, the
kink effective action is just a D=4 version of the D=10 super D2-brane, and the
kink-lump can indeed be identified as a BIon of this theory. 

Another result of this paper is a non-static but stationary
generalization of the kink-lump which we call a Q-kink-lump. It can be
viewed as a kink-lump boosted in the `hidden' fifth dimension.  In the
limit of infinite boost, to the speed of light, the Q-kink-lump
reduces to the Q-lump, so the Q-kink-lump is the generic 1/4 
supersymmetric soliton of the massive HK sigma models under
consideration. A boost of the D=5 supermembrane in the fifth dimension
corresponds to the inclusion of a constant background magnetic field
in the effective D=4 DBI action describing the kink domain wall. 
Using the methods of \cite{GGT}, we find the BIon solution in this
background and confirm its status as the effective description of the
Q-kink-lump. An interesting feature of this result is that the limit
of infinite boost, in which the Q-kink-lump becomes the Q-lump,
corresponds to a limit of infinite magnetic field in the DBI
theory.

Although we are concerned here with field theory solitons, most supersymmetric
field theories arise as effective theories in some superstring or M-theory
context, and their soliton solutions thereby acquire a superstring or M-theory
interpretation. The 1/2 supersymmetric 
kinks and Q-kinks of the models discussed here were
provided with several such interpretations in \cite{bergtown,LT}. Here we
shall show that the 1/4 supersymmetric kink-lump extends to a solution
of the M5-brane equations of motion, in a multi 
M-monopole background. As such, 
it provides an example of a calibrated M5-brane preserving 1/16 of the
supersymmetry of the M-theory vacuum. A similar result holds for the
Q-kink-lump (and hence the Q-lump) with the difference that the solution is
time dependent. It is thus a generalization of a calibration, of a type first
discussed in \cite{GLW}, that could be called a `Q-calibration'.

We shall begin with a discussion of the sigma model field theories and their
solitons, including the kink-lump and the Q-kink-lump, and their properties.
We then discuss the effective description of the kink-lump in
terms of a D=4 DBI action for a sigma-model D2-brane, and
show that the Q-kink-lump can then be found by considering the DBI action in a
constant background magnetic field. We then show how all these 1/4
supersymmetric solitons determine supersymmetric minimal energy configurations
of the M5-brane in a multi M-monopole background. We conclude with a discussion
of some other issues.

\section{Kinky Lumps}

The sigma models of relevance here have as their target 
space a HK 4-manifold of the type described above. 
The simplest choice of the harmonic function $U$ that
serves our purposes is 
\be
U = a + {1\over2}\left[{1\over |{\bf X}- {\bf n}|} + 
{1\over |{\bf X}+ {\bf n}|}\right]\, ,
\ee
where ${\bf n}$ is a unit 3-vector and $a$ a constant. The function $U$
is singular at the two `centres' ${\bf X}=\pm {\bf n}$, but this is 
a coordinate
singularity of the metric if $\varphi$ is periodically identified with period
$2\pi$. When $a=0$ we have the Eguchi-Hanson metric. For $a=1$ we have the
asymptotically flat metric transverse to two M-monopoles.  
In either case, the metric is HK with the triplet of K\"ahler 2-forms
\be
\bfO = (d\varphi + d{\bf X}\cdot {\bf A}) d{\bf X} - 
{1\over2}U\, d{\bf X}\times
d{\bf X}\, ,
\ee
the wedge product of forms being implicit.

The 2-centre metric (and, more generally, any multi-centre metric with colinear
centres) has an additional Killing vector field generating rotations about the
${\bf n}$ axis. This Killing vector field is holomorphic with respect to the
complex structure $I$ associated to the K\"ahler 2-form $\Omega = {\bf n}\cdot
\bfO$. The 3-vector ${\bf X}$ of $SO(3)$ can be decomposed into the singlet
$X={\bf n}\cdot {\bf X}$ and a doublet under the $SO(2)$ subgroup that fixes
${\bf n}$. The HK sigma-model can then be consistently truncated to a K\"ahler
sigma model by keeping only the singlet fields $(\varphi,X)$. Because
the truncation is consistent any solution of the reduced K\"ahler 
sigma-model equations will solve the full HK sigma-model equations. 
The metric on this 2-dimensional K\"ahler subspace of the target space is
\be\label{2sphere}
ds^2(K_2) = UdX^2 + U^{-1}d\varphi^2
\ee
where, for $|X|\le1$,
\be
U= a  + {1\over 1-X^2}\, .
\ee
The K\"ahler 2-form is $\Omega = {\bf n}\cdot \bfO$ and, since one can choose
${\bf A}$ such that ${\bf n}\cdot {\bf A}=0$ \cite{PT}, we have
\be
\Omega = d\varphi \wedge dX\, ,
\ee
which is the volume form on the 2-sphere. The lump charge $L$ is the
integral of the pull-back of $\Omega$ so its minimum value is $4\pi$,
the area of the two sphere. This is the tension of the singular
lump-string. 

Although D=5 is the maximal dimension in which we may have a massive
supersymmetric sigma model (a point that we return to in the concluding
discussion) it will be sufficient for our purposes to  consider a D=4 N=2 model
with Lagrangian density (\ref{lag}). For simplicity we shall set
$\mu=1$ and $g=1$. After the truncation to the N=1 
supersymmetric K\"ahler sigma
model described above, this yields the energy density
\be\label{linearized}
{\cal E} = {1\over2}\left[ U(\dot X^2  + |\bfn X|^2) + 
U^{-1}(\dot\varphi^2 + |\bfn\varphi|^2 + 1)\right]\, ,
\ee
which can be rewritten as
\bea
{\cal E} &=& {1\over2}\left[U\dot X^2 + U^{-1}(\nabla_1
\varphi)^2\right] 
+ {1\over2}\left[U^{-1}(\dot\varphi - v)^2 + U(\nabla_1 X \mp
\sqrt{1-v^2}\, U^{-1})^2\right] \nn
&&\ +\ {1\over2}\left[U^{-1}(\nabla_2\varphi - \sigma U\nabla_3 X)^2 +
U^{-1}(\nabla_3\varphi + \sigma U\nabla_2 X)^2\right]\nn
&&\ + \ v U^{-1}\dot\varphi \pm \sqrt{1-v^2}\,  \nabla_1 X 
+ \sigma (\bfn \varphi \times \bfn X)_1 
\eea
for constant $v$, with $|v|\le1$, and $\sigma= \pm 1$. Noting that 
\be
Q= \int\! d^3 x\,  U^{-1}\dot\varphi 
\ee
is a Noether charge (associated with the triholomorphic
isometry of the original HK target space metric) we see that the above
expression for the energy density implies (by appropriate choice of $v$)
the following (formal) bound on the total energy $E$:
\be
E \ge \sqrt{Q^2 + K^2} + |L|\, ,
\ee
where $K$ and $L$ are the topological kink and lump charges
\be
K= \int d^3x\, (\bfn X)_1 \, , 
\qquad L = \int d^3x \, (\bfn \varphi \times \bfn X)_1\, .
\ee
Note that $L$ is the pullback to the $23$-plane of the K\"ahler 2-form
$\Omega$. The bound is saturated when 
\be
\dot X=0\, ,\qquad\dot\varphi = v = {Q\over \sqrt{Q^2+K^2}}
\ee
and 
\be\label{kink}
\nabla_1 \varphi =0\, ,\qquad \nabla_1 X = \pm (\sqrt{1-v^2})\, U^{-1}
\ee
and
\be\label{lump}
\nabla_2\varphi = \sigma U\nabla_3 X\, ,\qquad 
\nabla_3\varphi = -\sigma U\nabla_2 X\, .
\ee

To solve these equations it will prove convenient to set 
\be
X= \pm \tanh u\, ,\qquad  \varphi=vt +\psi\, ,
\ee
for time independent $u, \psi$.
The equation for $X$ in (\ref{kink}) then becomes
\be
\partial_1 \left( u + a \tanh u\right) = \sqrt{1-v^2}\, .
\ee
Let us also set
\be
x^1=x,\qquad x^2 \pm ix^3 =z\, .
\ee
The function $u(x,z)$ is then given implicitly by 
\be
u + a \tanh u = \sqrt{1-v^2}\, x + \sigma\log w(z,\bar z)\, ,
\ee
and the two real equations (\ref{lump}) are now equivalent to the single 
complex equation
\be
\bar\partial \left(\psi + i\log w\right) =0\, ,
\ee
where $\bar\partial$ indicates a partial derivative with respect to $\bar z$.
Equivalently, 
\be\label{holo}
we^{-i\psi} = Z(z)
\ee
for arbitrary holomorphic function $Z$.

We have now found an implicit, but general, solution of the equations
(\ref{kink}) and (\ref{lump}). For $a=0$ 
the solution can be given explicitly. Choosing the upper sign 
and $\sigma=1$ we have 
\be
X = \tanh \left[ \sqrt{1-v^2}\, x + \log w \right]
\ee
with $\psi= -\arg Z$. For constant $Z$ both $\psi$ and $w$ are
constant and we recover the Q-kink solution of \cite{AT}. Other choices of
$Z(z)$ yield new solutions. For example, we could have $Z=\lambda/z$ for
arbitrary complex constant $\lambda$. 
Consider, for simplicity,
\be
Z(z) = {1\over z}\, .
\ee
In this case $\psi = \arg z$ and
\be
X = \tanh \left[ \sqrt{1-v^2}\, x - \log |z| \right]\, .
\ee
For fixed $z$ we have a kink solution but for fixed $x$ we have a sigma-model
lump solution. This can be seen, for example, by noting that
$X\rightarrow 1$ as $z\rightarrow 0$ and $X\rightarrow -1$ as 
$z\rightarrow \infty$. For fixed $x$ the sigma model lump has scale
size $\exp(\sqrt{1-v^2}x)$. This is the simplest Q-kink-lump solution. 
The static kink-lump is found by setting $v=0$ while the
Q-lump is obtained by setting $v=1$.

We now turn to a determination of the fraction of supersymmetry preserved by
the kink-lump, Q-kink-lump and Q-lump. A formula for supersymmetric 
configurations of D=6 sigma-models with $4k$-dimensional toric HK target spaces
was obtained in \cite{GTT}. Specializing to the $k=1$ case we conclude that
supersymmetric configurations of D=6 sigma models are those for which the
equation  
\be\label{equation}
\left[\Gamma^m \bft \cdot  \partial_m {\bf X}  + iU^{-1} \Gamma^m
{\cal D}_m \varphi\right] \epsilon =0
\ee
admits solutions for non-zero $D=6$ $Sp_1$-Majorana-Weyl spinor $\epsilon$,
where $\bft$ are the Pauli matrices and $\Gamma^m$ ($m=0,1,2,3,4,5$) 
are the D=6
Dirac matrices. To apply this formula we note that the massive D=4 sigma model
discussed here is obtained from the D=6 model by setting
\be
\partial_4 {\bf X}= \partial_5 {\bf X} = 0\,, \qquad 
\partial_4 \varphi = 1\,, \qquad \partial_5 \varphi =0\, . 
\ee
Given also that
\be
{\bf X} = X\, {\bf n}\,, \qquad  \dot X=0\, ,
\ee
the supersymmetry condition becomes
\be
\left[ U (\bft\cdot {\bf n}) (\bfG \cdot \bfn X) + i(\bfG \cdot \bfn \varphi)
+ i\Gamma^0 \dot\varphi + i\Gamma^4\right]\epsilon =0
\ee
where $\bfG=(\Gamma^1,\Gamma^2,\Gamma^3)$. For the Q-kink-lump this
yields
\be
\left[ 1-(v\Gamma^{04} \mp i\sqrt{1-v^2}\, (\bft\cdot{\bf n})
\Gamma^{14})\right]\epsilon + \Gamma^4(\bfG \cdot \bfn\varphi)\left[1-
i\sigma  (\bft\cdot{\bf n})\Gamma^{23}\right]\epsilon =0\, .
\ee
When $\bfn \varphi$ vanishes we have
\be
\left(v\Gamma^{04} \mp i\sqrt{1-v^2} \Gamma^{14}\bft \cdot{\bf
n}\right)\epsilon=\epsilon \, ,
\ee
which confirms the 1/2 supersymmetry of the kink and Q-kink. When $\bfn
\varphi$ is non-zero we have, in addition, that
\be
i\Gamma^{23}(\bft\cdot{\bf n}) \epsilon = \epsilon\, .
\ee
The combined conditions imply 1/4 supersymmetry, for any $v$. We conclude that
the kink-lump, Q-lump and Q-kink-lump are all 1/4 supersymmetric. 

\section{Energetics}

We shall now set $a=0$ for simplicity, and again choose $\sigma=1$.
Then, when the Q-kink-lump solution is used in the expression for
the energy density, one finds that
\be\label{enden}
{\cal E} = {4e^{2y}\over \left(1+e^{2y}|Z|^2\right)^2}\, 
\left[|Z|^2 + |Z^\prime|^2\right]
\ee
where we have set
\be
y= \sqrt{1-v^2}\, x
\ee 
for convenience. If we integrate the energy density over $x$ we find the 
energy density on the domain wall to be
\be\label{wall}
{\cal E}_{wall} = 2\gamma \left(1 + |Z^\prime|^2/|Z|^2\right)\, ,
\ee
where
\be
\gamma={1\over \sqrt{1-v^2}}
\ee
By taking $Z$ to be constant we see that the wall's surface tension is 
$2\gamma$.

For the moment we postpone the analysis of ${\cal E}_{wall}$ 
for non-constant $Z$ and
return to the unintegrated formula (\ref{enden}). 
For a  general Q-kink-lump solution with 
\be\label{genlump}
Z(z)= \sum_i {\lambda_i \over z- z_i} 
\ee
we have, in the limit of large $r=|z|$,
\be
{\cal E} \sim {4 e^{2y}|Z|^2\over \left(1+e^{2y}|Z|^2\right)^2}\ .
\ee
This has a maximum when $e^y|Z|=1$, so we can take this as the
surface to which the domain wall is asymptotic at large $r=|z|$. 
This implies that for large $r$ the domain wall is asymptotic to the surface
\be\label{logbend}
y = \log r\, ,
\ee
unless $\sum_i\lambda_i=0$, in which case $y \sim \log r^2$. 

To proceed we shall now focus on the one-lump case with $Z=1/z$.
In this case the  energy density is
\be
{\cal E}(r,y) = {4(1+r^2)e^{2y}\over \left(e^{2y} + r^2\right)^2}\, .
\ee
This function is plotted in Figure 1 for a range of the independent variables $r$
and $y$. 
\begin{figure}
%\vspace{2mm}
\begin{center}
\leavevmode
\hbox{%
\epsfysize=5in
\epsffile{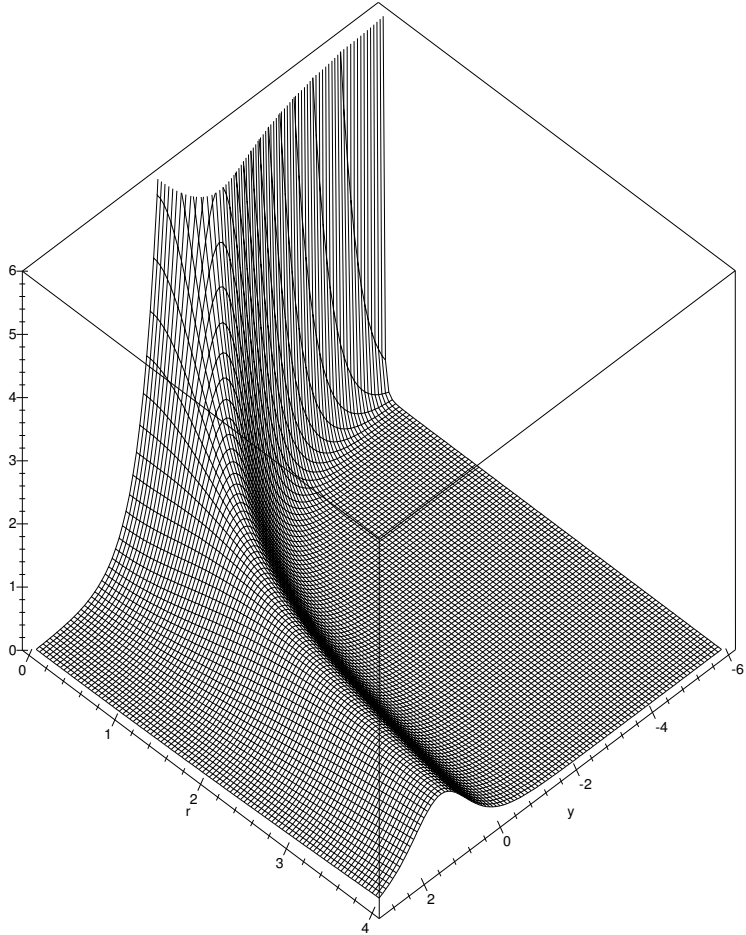}}
\end{center}
 \caption[f2]{A plot of the energy function ${\cal E}(r,y)$.}
\end{figure}  
The function ${\cal E}(r,y)$ has no extrema 
(except in the $r\rightarrow \infty$
limit discussed above) but some understanding of the solution near 
$r=0$ can be had by considering the extrema of the cross-sectional 
energy density ${\cal E}(r)$ at fixed $y$. As already
noted, the solution for fixed $y$ is a lump that interpolates between $X=1$ at
$r=0$ and $X=1$ at $r=\infty$. The function ${\cal E}(r)$ has an
extremum at $r=0$ and, if
\be
y > y_* \equiv \log\sqrt{2}\, ,
\ee
at 
\be
r = r_*(y) \equiv \sqrt{e^{2y} - 2}\, .
\ee
When $y>y_*$ the extremum at $r=0$ is a minimum and the extremum at $r=r_*$
is a maximum. The cross-sectional energy density is therefore ring-shaped for
sufficiently large $y$. The radius of the ring shrinks as $y$
increases; this is the advertized `spike', which is essentially hollow
for $y >y_*$. The radius of the ring shrinks to zero at $y=y_*$ and
for $y<y_*$ the only extremum of ${\cal E}(r)$ is a maximum at $r=0$. 
The solution remains non-singular in
that the energy density remains everywhere smooth and finite, but the
cross-sectional lump is no-longer ring-shaped. The hollow `spike' for $y>y_*$
is `filled in' for $y<y_*$, as one might expect from the fact that the domain
wall has a finite width, of order one in our units. This behaviour is
shown in Figure 2 in which ${\cal E}$ is plotted as a function of $r$
for values of $y>y_*$, $y=y_*$ and $y<y_*$.
\begin{figure}
%\vspace{2mm}
\begin{center}
\leavevmode
\hbox{%
\epsfysize=2in
\epsffile{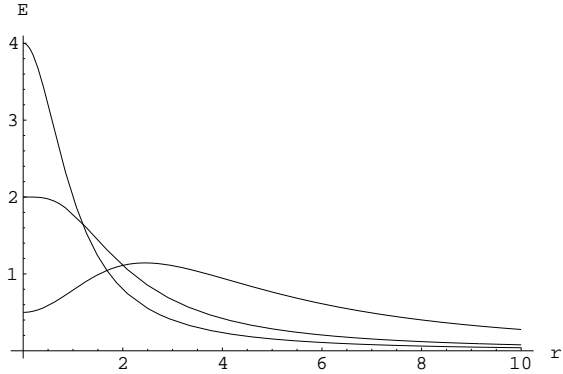}}
\end{center}
 \caption[f2]{A plot of ${\cal E}(r,y)$ for three fixed values
of y with $y>y_*$, $y=y_*$ and $y<y_*$, specified by the values of
${\cal E}$ at $r=0$ given by 1/2, 2 and 4, respectively.}
\end{figure}  
A natural interpretation of
this result is that the string is actually attached to the wall at the point at
which $e^{2y}=2$,  the wall being deformed by the string's tension just so 
as to meet the string endpoint at this distance. 

For $y\gg y_*$ it is natural to interpret $r_*\sim e^y$ as the size of the
cross-sectional lump. This implies that we have a domain wall with a shape
that is again given by $y \sim \log r$, consistent with the asymptotic
behaviour as $r\rightarrow \infty$ that we found earlier. But we also wish to
determine the shape for $y\ll y_*$. One way to do this would be to determine
the size of the cross-sectional lump as a function of $y$. Since the energy
density is centred at $r=0$ for $y\ll y_*$ the size is not related to the
position of the maximum of ${\cal E}$ for fixed $y$, as it is for $y\gg y_*$.
Naively, we might define the size as 
\be
\langle r\rangle \equiv \int_0^\infty (2\pi r)dr \, r {\cal E}(r)\, ,
\ee
but this integral diverges for the simple lump solution with $Z=\lambda/z$.
In fact the integral of ${\cal E}$ also diverges. Both divergences may be
removed by considering a multi-lump solution of the form (\ref{genlump}) with
$\sum_i\lambda_i=0$ but the value of $\langle r\rangle$ is then more
naturally interpreted as the mean distance between the constituent lumps (as
discussed for the Q-lump in \cite{Abraham}). Since the size of an individual
lump is really determined by the energy density for small $r$ we shall proceed 
by first noting that
\be
{\cal E} \approx {4e^{2y}\over \left(e^{2y} + r^2\right)^2}\, .
\ee 
for $r\ll 1$. This has a finite integral over the $z$-plane, and $\langle
r\rangle$ is also finite. In fact
\be\label{const}
\langle r\rangle = const. \times e^y\, .
\ee
The constant depends on the particular `regularization' used. 
Its value will not be important to us but one may note that we could have
defined the shape of the spike in terms of the surface on 
which ${\cal E}(y)$ is
a maximum for fixed $r$. This surface is $r=e^y$, so we 
have agreement with the 
result of considering ${\cal E}(r)$ for fixed $y$ if we set the constant in 
(\ref{const}) to unity.  

The final conclusion of this analysis is that the shape
of the spike for $y\ll y_*$ is given by 
\be
y = \log r \, ,
\ee
just as it was for $y\gg y_*$. A cut off at a distance $\delta$ from this
singularity therefore corresponds to a distance $\ell$ from the wall with $\ell$
related to $\delta$ by
\be
\label{Lerel}
-\log\delta = \sqrt{1-v^2}\, \ell + const\, .
\ee

We now return to the formula (\ref{wall}) for the energy density on the wall. 
Let us again take $Z=1/z$ and integrate over the $z$-plane, with IR cutoff 
at $r=R$ and UV cutoff at $r=\delta$. We find that
\be\label{toberec}
E= 2\gamma\left(\pi R^2 + 2\pi \log R\right) + 4\pi \ell + const. + 
\dots 
\ee
where we have used the relation (\ref{Lerel}) to convert the $\delta$ 
dependence to a dependence on $\ell$, and the terms omitted vanish in the
limit of $\delta\rightarrow 0$.  The $R^2$ term can be considered as the
vacuum energy of the domain wall. The $\log R$ term is the expected IR
divergent energy of a global vortex in D=3. The term linear in $\ell$ can be
interpreted as the energy in a string of length $\ell$ and tension
\be
T_{string}= 4\pi\, .
\ee
This is precisely the tension of the singular lump-string, so the
natural interpretation is that the kink-lump provides a D=4 spacetime
description of a normally-singular lump-string ending on a D=4 Q-kink domain
wall.  

Note that all of the above discussion applies for any value of the parameter 
$v<1$, in particular for $v=0$, which yields the static kink-lump solution. 
We now turn to the limiting case of $v=1$. In this limit the 
tube-like mid-section of the lump-string gets stretched out, with the wall 
itself, and the `solid spike' region, being pushed off to infinity. 
We then have an infinite straight Q-lump-string, i.e. a string 
with a Q-lump core and cross-sectional energy density 
\be
{\cal E} = {4(|Z|^2 + |Z^\prime|^2)\over (1+ |Z|^2)^2}\, .
\ee
As noted by Abraham, the integrated cross-sectional energy, i.e. 
the Q-lump string tension, is infinite for a single lump, but is 
finite for a multi-lump solution with $\sum_i\lambda_i=0$. For
example, for any complex constant $a$ the choice
\be
Z= {1\over z-a} - {1\over z+a}
\ee
leads to a non-singular and finite energy charge-two Q-lump solution.
We refer to \cite{Abraham} for detailed properties of multi Q-lumps, 
but a plot of the energy density for the above two Q-lump solution 
is shown for $a=1/2$ in Figure 3.

\begin{figure}
%\vspace{2mm}
\begin{center}
\leavevmode
\hbox{%
\epsfysize=3in
\epsffile{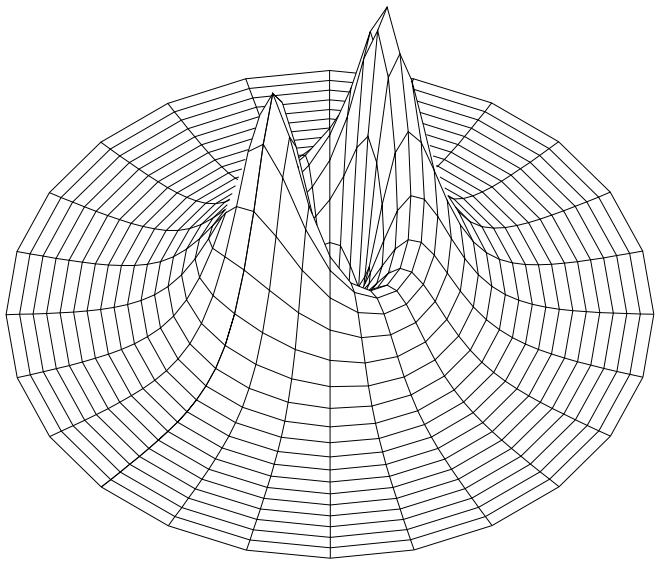}}
\end{center}
 \caption[f2]{ A plot of the energy density ${\cal E}(z)$ for 
a charge two Q-lump.}
\end{figure}

\section{Effective D-brane description}

As mentioned in the introduction, the new kink-lump solution of massive 
HK sigma models that we have found and studied here is similar in some 
respects to the BIon solution of the DBI field equations describing 
the fluctuations of a IIA superstring theory D2-brane \cite{CM}. However, the
proper analogy of the kink-lump in this context would be to a IIA
{\sl supergravity} solution in which a string ends on a D2-brane, 
because only in this case would the D2-brane be part of the solution.
In this sense, the proper sigma-model analogue of the BIon is found by
asking whether the 1/4 supersymmetric kink-lump solution can be be 
understood as a 1/2 supersymmetric solution on the effective D=3 field 
theory governing the fluctuations of the kink domain wall. Indeed, it can be
understood this way, as we now describe. 

The general static kink solution is given implicitly by
\be
X = \pm \tanh [(x-x_0) \mp aX]\, \qquad \varphi = \varphi_0
\ee
where $x_0$ and $\varphi_0$ are two real collective coordinates, with
$\varphi_0 \sim \varphi + 2\pi$. Identification of the collective coordinates
as the coordinates of the space transverse to an infinite planar membrane, and
the fact that the kink solution preserves 1/2 of the eight sigma-model
supersymmetries implies that the kink has an effective description as a 
supermembrane in a D=5 $\bE^{1,3}\times S^1$ spacetime \cite{AT}. To see this
we allow the collective coordinates to become smooth functions of the 
worldvolume coordinates $\xi^i$ ($i=0,1,2$) to arrive at the worldvolume fields
\be
\phi(\xi) \equiv x_0(\xi)\, ,\qquad \sigma(\xi) \equiv \varphi_0(\xi)\, ,
\ee
which may be identified with the physical (transverse) 
boson fields of the supermembrane
in the gauge in which three worldvolume fields taking 
values in $\bE^{(1,2)}$ are
identified with the coordinates of an $\bE^{(1,2)}$ subset 
of the D=5 spacetime.
The physical worldvolume fields thus determine the position of a membrane in
the $\bE^{(1,3)}\times S^1$ spacetime. The symmetries of the kink solution then
imply that the low energy effective action for these fields is that of the 
D=5 supermembrane \cite{BST}. As the kink domain wall tension equals 2 in our
mass units, the bosonic action is
\be
I = -2\int d^3\xi \sqrt{-\det (g_{ij} + \partial_i\sigma\partial_j
\sigma)}
\ee
where $g_{ij}$ is the metric induced from the D=4 Minkowski metric. In a
physical gauge it is given by
\be
g_{ij} = \eta_{ij} + \partial_i \phi\partial_j \phi\, .
\ee

Because $\sigma$ is periodically identified, $d\sigma$ is the dual 
of a $U(1)$ worldvolume 2-form field strength. The dual field theory is 
just the D=4 DBI action (for the same reasons that the D2-brane action
is dual to the D=11 supermembrane action in a $\bE^{(1,9)}\times S^1$
background \cite{pkt}). The bosonic action is 
\be\label{dbi}
I = -2\int d^3\xi \sqrt{-\det (g_{ij} + F_{ij})} 
\ee
where the on-shell relation of the BI two-form field-strength
$F$ to $\sigma$ is given by
\be
\sqrt{-\det g} g^{ij}\partial_j \sigma = 
{1\over2}\sqrt{1+ (\partial\sigma)^2} \, 
\varepsilon^{ijk} F_{jk}
\ee
where $(\partial\sigma)^2 = g^{ij}\partial_i \sigma\partial_j\sigma$.
Note that the solution of the supermembrane
equations with $d\phi=0$ and $d\sigma = vdt$ corresponds to
a solution of the DBI equations with $d\phi=0$ and
\be\label{fsig}
F= - {v\over \sqrt{1-v^2}}\, d\xi^1\wedge d\xi^2\, ,
\ee
so that $B\equiv F_{12}$ is a constant related to $v$ by 
\be\label{Bvee}
\sqrt{1+ B^2} = \gamma(v)\, .
\ee

The above discussion for the static kink domain walls can 
be generalised to the stationary solutions by expanding the
above DBI action
about a non-zero but constant magnetic background field $B$ given by
(\ref{fsig}). We begin with a formula of \cite{GGT} for the physical
gauge DBI energy density ${\cal H}$. For static 2-brane
configurations this formula is 
\be
{\cal H}^2 = 4\left[(1+|{\bf E}|^2) (1+ B^2) + ({\bf E}\cdot \bfn \phi)^2 +
|\bfn \phi|^2\right]
\ee
where ${\bf E}$ is the electric field. Assuming that $B$ is constant,
and related to the constant $\gamma(v)$ by (\ref{Bvee}), we may
rewrite this as
\be
{\cal H}^2 = 4(\gamma \pm  {\bf E}\cdot\bfn \phi)^2 + 
4|\gamma{\bf E} \mp \bfn \phi |^2 \, .
\ee
Following the argument of \cite{GGT} we deduce the bound
\be
\int d^2\sigma\left[{\cal H} -2\gamma\right] \ge  
2\big|\int d^2\sigma {\bf E}\cdot \bfn \phi\big|
\ee
with equality when
\be
\gamma {\bf E} =  \pm \bfn \phi\, .
\ee
This implies that $\phi$ is harmonic and we may choose the unit point charge
solution 
\be\label{exx}
\phi = \gamma\, \log r\, ,
\ee
for which ${\bf E} = {\bf e}_r/r$ where ${\bf e}_r$ is a unit vector directed
radially outwards\footnote{For $v=0$ this solution corresponds 
to the D=5 supermembrane configuration 
$\phi + i\sigma = - \log \zeta$,
where $(t,\zeta)$ ($\zeta$ complex) parameterize the membrane's worldvolume.
For $v=0$ similar solutions are well-known in string theory, e.g. as a
D4-brane ending on an NS5-brane \cite{witten}. In this case we have a
sigma-model lump-string ending on a sigma-model kink-membrane.}.

To perform the integral of ${\bf E}\cdot \bfn \phi$ we introduce an
IR cutoff at $r=R$ and a UV cutoff at $r=\delta$. 
The total energy of the point charge solution is then
\be
H= 2\gamma \left[\pi R^2 -\pi \delta^2\right] + 2\bigg|\phi(R)\oint_{r=R} {\bf
dS}\cdot {\bf E} + \phi(\delta) \oint_{r=\delta} {\bf dS}\cdot {\bf E}\bigg|
\ee
where ${\bf dS}$ is an outward pointing line element on a curve enclosing the
origin. The integrals are easily done, with the result that 
\be
H = 2\gamma \pi R^2 + 4\pi \left[\phi(R)-\phi(\delta) \right] + \dots\, ,
\ee
where the terms neglected vanish in the limit that $\delta\rightarrow 0$. 
Using the formula (\ref{exx}) and the fact that $-\phi(\delta) =\ell$, where
$\ell$ is distance from the 2-brane, we find that
\be
H= 2\gamma\left[ \pi R^2 + 2\pi \log R\right] + 4\pi \ell + \dots
\ee
in complete agreement with the formula (\ref{toberec}) for the energy of a
Q-kink-lump. The agreement confirms that we have correctly
identified the DBI action as the effective action of the kink domain wall and
that we have correctly identified the 1/2 supersymmetric BIon solution of the
latter with the 1/4 supersymmetric kink-lump solution of the sigma
model.

\section{M-theoretic interpretation}

We shall provide the 1/4 supersymmetric sigma model solitons discussed here 
with an M-theoretic interpretation by showing that they yield solutions of 
the M5-brane equations of motion in a D=11 supergravity background with 
vanishing 4-form field strength and 11-metric
\be\label{background}
ds^2= -dT^2 + ds^2(\bE^4) + dS^2  + ds^2(HK_4) + dZ^2\, .
\ee
We take the Killing vector field
\be
\ell = {\partial\over\partial S}
\ee
to generate a $U(1)$ isometry, and $HK_4$ to be 
 a multi-centre 4-metric of the type considered above. This has an M-theory 
interpretation as the metric produced by M-monopoles (situated at the 
centres of the 4-metric). 

We now consider a 5-brane in this background. As the background breaks
half the supersymmetry of the M-theory vacuum, the field theory on the
M5-brane has a D=6 (1,0) supersymmetry and the field content splits
into a tensor multiplet and a hypermultiplet. There is a consistent
truncation to the hypermultiplet sector, and we shall perform this
truncation. For an appropriate choice of the M5-brane vacuum the low
energy field theory is then a massless D=6 sigma model with a
multi-centre HK 4-manifold as its target space. The massive D=5 sigma
model is then found as the effective field theory on an M5-brane
wrapped on a particular combination of $S^1$ cycles in the background,
as described for the M2-brane in \cite{bergtown}. A further trivial
double-dimensional reduction yields the massive D=4 sigma model
discussed above. 

To specify the needed M5-brane configuration we begin by taking 
$({\bf Y},W)$ to be the $\bE^4$ coordinates and $X^I$ ($I=1,2,3,4$) the $HK_4$
coordinates. A 5-brane configuration is then specified by giving the 11
spacetime coordinates $X^M=(T,{\bf Y},S,W,X^I,Z)$ as functions of the six 
worldvolume coordinates $(t,{\bf y},s,w)$. Six of these functions may be chosen
so as to fix the worldvolume diffeomorphism invariance of the fivebrane action.
We shall make the `physical gauge' choice
\be
T=t, \qquad {\bf Y}={\bf y}\, ,\qquad S=s\, ,\qquad W=w\, .
\ee
This leaves $(X^I,Z)$ as the physical worldvolume fields, specifying the
deformation of the 5-brane in the transverse 5-space from the vacuum
configuration in which $Z$ and $X^I$ are constant. We will set $Z$ to
a constant as its fluctuations belong to the fields of the tensor multiplet 
that we are discarding. We are left with the four worldvolume 
scalar fields $X^I$ (and
their superpartners). In principle, these fields are functions of all six
worldvolume coordinates, but we will impose invariance under shifts of $w$. 
This leaves us with the D=5 fields $X^I(\xi)$ (and their superpartners) where
\be
\xi^\mu = (t,{\bf y},s)\, .
\ee
Eventually we will impose the constraint
\be\label{SSred}
\partial_s X^I \equiv \partial_4 X^I =\zeta^I
\ee
where $\zeta$ is the tri-holomorphic KVF of the HK 4-metric, thus reducing the
effective field theory to a massive D=4 supersymmetric sigma model. 
What we will now show is that the Q-kink-lump solution of this effective
field theory defines an M5-brane configuration preserving 1/4 supersymmetry.

The number of supersymmetries preserved by a given 
M5-brane configuration is the
dimension of the space of solutions for the constant spinor $\epsilon$ to the
condition \cite{BBS,bergtown}
\be\label{susycon}
\Gamma \epsilon = \epsilon
\ee
where $\epsilon$ is a Killing spinor of the background, and $\Gamma$ is an
11-dimensional Dirac matrix function to be specified below. In the 
present case,
there are 16 linearly independent Killing spinors, satisfying
\be\label{con1}
\Gamma_{1234}\epsilon=\epsilon\, .
\ee
The Killing spinors have the form $\epsilon= f\epsilon_0$ for a universal
function $f$ and constant spinor $\epsilon_0$. As $f$ cancels from 
(\ref{susycon}) we may replace $\epsilon$ by $\epsilon_0$ in this equation;
having done so we may then drop the suffix on $\epsilon$ to arrive back at
(\ref{susycon}) but with $\epsilon$ now taken to be 
a {\sl constant} spinor satisfying the
constraint (\ref{con1}). 

To specify $\Gamma$ we begin by taking the D=11 Dirac matrices to be
\be
\Gamma_M = (\gamma_\mu,\gamma_5,\Gamma_I, \gamma_*\Gamma_{1234})
\ee
where
\be
\gamma_* = \gamma_{01234}\gamma_5
\ee
Thus $\gamma_5^2=\gamma_*^2=1$, and the remaining non-zero anticommutators
are
\be
\{\gamma_\mu,\gamma_\nu\}= 2\eta_{\mu\nu}, \qquad  
\{\Gamma_I,\Gamma_J\}=2G_{IJ}\, .
\ee
The matrix $\Gamma$ for bosonic M5-brane
configurations with vanishing worldvolume 3-form field strength is
then
\be\label{gam}
(\sqrt{-\det g})\Gamma = {1\over 5!} 
\varepsilon^{\mu\nu\rho\lambda\sigma} \partial_\mu
X^M\partial_\nu X^N \partial_\rho X^P \partial_\lambda X^Q 
\partial_\sigma X^R\, \Gamma_{MNPQR}\gamma_5
\ee
where $g$ is the worldvolume 6-metric. In the physical gauge (and with $Z=0$) 
it is block diagonal with components ${\rm diag}(g_{\mu\nu},1)$, where
\be
g_{\mu\nu} = \eta_{\mu\nu} + \partial_\mu X^I\partial_\nu X^J G_{IJ}\, .
\ee
We can now rewrite the condition (\ref{susycon}) as
\be\label{gam2}
\left(\sqrt{-\det g}\right)\epsilon = \big(1 - \gamma^\mu \partial_\mu X^I
\Gamma_I -  {1\over2} \gamma^{\mu\nu}\partial_\mu X^I \partial_\nu X^J
\Gamma_{IJ} + \dots \big)\gamma_*\epsilon\, .
\ee
In principle, the right hand side includes terms up to $4$th order 
in $\partial X$ (recall that there are only 4 sigma-model fields $X^I$) 
but terms higher than second order vanish for a configuration such as the
kink-lump that depends on only two of the four sigma-model fields. 

Since the equations (\ref{kink}) and (\ref{lump}) are linear in $(\partial X)$,
the supersymmetry preservation condition (\ref{gam2}) 
must be satisfied order by order. At zeroth order we have 
\be\label{zeroth}
\gamma_*\e=\e,
\ee
which tells us that the vacuum state of the M5-brane is a $1/2$
supersymmetric M-theory configuration. The constraints (\ref{con1}) and
(\ref{zeroth}) preserve 1/4 of the 32 supersymmetries of the M-theory vacuum;
that is, they preserve 8 supersymmetries, which is the expected number for the
vacuum of a supersymmetric HK sigma model. At first order we have
\be\label{first}
\gamma^\mu\partial_\mu X^I \Gamma_I\epsilon =0\, .
\ee
Because the sigma-model is obtained by retaining the terms quadratic
in $\partial X$ in a series expansion of the 5-brane 
action, (\ref{first}) is equivalent to the field theory condition for
preservation of supersymmetry, as we shall verify below. 

The higher order terms in (\ref{gam2}) are now identities. The
analysis is similar to that of  \cite{bergtown}. We first note that 
(\ref{first}) implies
\be
\gamma^{\mu\nu}\partial_\mu X^I\partial_\nu X^J \Gamma_{IJ} = - \eta^{\mu\nu}
\partial_\mu X^I\partial_\nu X^J G_{IJ}\, ,
\ee
which in turn implies that (\ref{gam2}) is satisfied if, and only if,
\be
\det (\eta + G) = (1 + {1\over2} \tr G)^2\, .
\ee
That this is indeed satisfied follows from the fact that the rank of $G$ 
cannot exceed 2 because there are only two `active' fields ($X,\varphi$).

We now apply the above result to the Q-kink-lump to confirm that it preserves 
1/4 supersymmetry.  By use of the Q-kink-lump equations, (\ref{first}) can be
shown to be equivalent to  two further conditions on $\epsilon$. One is
\be
\gamma^{23} \Gamma \Gamma_\varphi \epsilon = \sigma\epsilon
\ee
where we have used $\Gamma_I= (\bfG,\Gamma_\varphi)$ and set 
$U^{-1}{\bf n}\cdot \bfG = \Gamma$, so that $\Gamma^2=\Gamma_\varphi^2=1$.
This is the `lump' condition which, by itself, preserves 1/2 of the 8 sigma
model supersymmetries. The other condition is
\be
\Gamma_v \epsilon=-\epsilon
\ee
where
\be
\Gamma_v \equiv v\gamma^{04}\pm \sqrt{1-v^2}\,
\gamma^{14}\Gamma\Gamma_\varphi\, .
\ee
Note that $\Gamma_v^2=1$ and $[\Gamma_v,\gamma^{23} \Gamma \Gamma_\varphi]=0$,
so this additional condition reduces the supersymmetry to 1/4 of the sigma
model vacuum. Note that this is true even if $v=1$, in which case the
Q-kink-lump reduces to the Q-lump. Thus, both the Q-lump and the Q-kink-lump
define M5-brane configurations preserving 1/16 of the supersymmetry of the
M-theory vacuum,  corresponding to 1/4 of the supersymmetry of the sigma model
vacuum. 

\section{Discussion}

We have seen that much of the physics of D-branes can appear in a purely 
field theoretic context. It is natural to ask whether the D-brane analogy 
can be stretched further. One obvious question is whether non-abelian symmetry 
enhancement occurs for coincident kink domain walls. 
The first point to appreciate here is that not every model
with  kink solutions will have static multi-kink solutions. In the simple
2-centre model considered here there are not even multi-kink 
configurations. To get multi-kink configurations one needs 
either (i) a {\sl multi-centre} target space 4-metric, or 
(ii) a higher-dimensional target space metric. In the first case,
a model with co-linear centres has obvious multi-kink configurations,
but no {\sl static} multi-kink solutions because the kinks repel (as will
be shown elsewhere \cite{GTTtwo}). This behaviour can also occur in string
theory \cite{string}, where it is attributable to 
non-abelian instanton effects in the D=3 N=2 SYM theory on the branes. This
suggests that a similar non-abelian gauge theory interpretation may be
possible for sigma-model D-branes. 

We chose to set $\mu=1$ and $g=1$ throughout most of the paper. If one
reinstates them one finds, for example, that the DBI action (\ref{dbi}) becomes
\be
I = -{2\mu\over g^2}\int d^3\xi \sqrt{-\det (g_{ij} + \mu^{-1}F_{ij})} 
\ee
and the wall and string tensions become 
\be
T_{wall} = 2\mu/g^2\, ,\qquad T_{string} = 4\pi/g^2\, . 
\ee
Recall that the Q-kink-lump was recovered by expanding about a
constant background magnetic field $B$, and the the Q-lump was obtained in
a limit corresponding to infinite $B$. Following \cite{SW} one can 
rescale $\mu$ and $g$ in this limit to end up with a non-commutative
D=3 gauge theory. This suggests that the sigma-model Q-lump may have
an alternative description as a non-commutative soliton. 

A difference between the D=5 supermembrane of relevance to sigma-model
D-branes and the M2-brane of relevance to string theory D-branes is
that that D=5 membrane can be viewed as an $S^1$ wrapped D=6 3-brane
(whereas the M2-brane has no analogous D=12 precursor). The D=4 D2-brane is 
thus a 3-brane in a D=6 spacetime of the form $\bE^{(1,3)}\times T^2$ that has
been wrapped on a homology cycles of the 2-torus. This is to be expected from
the fact that D=5 is the maximal dimension for massive HK sigma-models while
we considered only the D=4 models. The kink is a 3-brane of the D=5 massive
sigma-model and a sigma model lump is a 2-brane. The kink lump solution thus
lifts to a solution of the D=5 model representing a 2-brane with a string
boundary on the 3-brane. 

Finally, we note that the results of section 5
can be stated in terms of calibrations. Recall that 
the lump solution of the massless sigma model corresponds to
a K\"ahler calibrated two surface in four dimensions \cite{GPT}.
We now have a similar interpretation of the
kink-lump of the massive sigma model as a K\"ahler calibrated 
4-surface in six dimensions. The Q-kink-lump, on the other hand, 
is not a calibrated 4-surface, strictly speaking, because 
it is time dependent. This kind of `time-dependent calibration'
has been discussed in \cite{GLW} and we suggest the terminology
`Q-calibration'. As we have seen, Q-calibrations are 
{\sl stationary}, but not necessarily static,  minimal energy
surfaces. The Q-kink-lump is therefore a K\"ahler Q-calibrated 
4-surface in six dimensions. It reduces for $v=1$ to the Q-lump, 
which is a K\"ahler Q-calibrated two surface in four dimensions.

\vskip 0.5cm 
\noindent {\bf Acknowledgements}:
PKT would like to thank Taichiro Kugo and Nobuyoshi Ohta for helpful
discussions, and the ESI (Vienna), the YI (Kyoto) and the Kyoto University
Department of Physics for their hospitality. DT is supported by an EPSRC
fellowship. RP thanks Trinity College, Cambridge for financial support.
All authors are supported in part by PPARC through SPG $\sharp$ 613.

\end{document}